\documentclass[a4paper, 11pt, times, authoryear]{elsarticle}
\biboptions{sort&compress}

\usepackage{amsmath}
\usepackage{siunitx}
\usepackage{hyperref}
\usepackage{tabularx}

\usepackage{geometry}
\usepackage[colorinlistoftodos, textwidth=2.cm, textsize=footnotesize,
             shadow, backgroundcolor=green, linecolor=green]{todonotes}  

\makeatletter
\newcommand*{\rom}[1]{\expandafter\@slowromancap\romannumeral #1@}
\makeatother\newcommand*\diff{\mathop{}\!\mathrm{d}}
\usepackage{upgreek}
\usepackage{graphicx}
\usepackage{subfigure}
\usepackage{color}
\usepackage{caption}
\usepackage{lipsum}
\usepackage{tabularx}
\usepackage{booktabs}

\usepackage{color}
\usepackage[normalem]{ulem} 
\newcommand{\Replace}[2]{\bgroup\noindent\textcolor{red}{\xout{#1}#2}\egroup\ignorespacesafterend}
\newcommand{\Delete} [1]{\bgroup\noindent\textcolor{red}{\xout{#1}}\egroup\ignorespacesafterend}
\newcommand{\Insert} [1]{\bgroup\noindent\textcolor{red}{#1}\egroup\ignorespacesafterend}
\newcommand{\Comment}[1]{\definecolor{Mygray}{gray}{0.50}\bgroup\color{Mygray}\noindent#1\egroup\ignorespacesafterend}
\newcommand \Ronghai [1]{\bgroup\noindent[\textcolor{blue}{\textbf{Ronghai}: #1}]\egroup\ignorespacesafterend}
\newcommand \Stefan  [1]{\bgroup\noindent[\textcolor{blue}{\textbf{Stefan}: #1}]\egroup\ignorespacesafterend}
\newcommand \Michael  [1]{\bgroup\noindent[\textcolor{red}{\textbf{Michael}: #1}]\egroup\ignorespacesafterend}

\DeclareMathAlphabet{\Ibb}{U}{msb}{m}{n}
\newcommand   {\IC}{{\ensuremath{\Ibb C}}}
\newcommand \ggp {{\ensuremath{{{\gamma}/{\gamma}^{\prime}}}}}
\newcommand \gp  {{\ensuremath{{\gamma}^{\prime}}}}
\newcommand \g   {{\ensuremath{\gamma}}}
\newcommand{\matone}{\ensuremath{\text{\textup{\textbf{I}}}}}

 \newcommand  {\half     }{{\textstyle\frac{1}{2}}}
 \renewcommand{\ldots}{\mathinner{\mkern-1mu\ldotp\mkern-2mu\ldotp\mkern-2mu\ldotp\mkern-1mu}}
 \newcommand{\allphi}{\phi_1,\ldots,\phi_4}  
 
 \newcommand{\Bb}{{\boldsymbol{\mathnormal b}}}

 \newcommand{\Be}{{\boldsymbol{\mathnormal e}}}

 \newcommand{\Bn}{{\boldsymbol{\mathnormal n}}}
 \newcommand{\Bu}{{\boldsymbol{\mathnormal u}}}
 \newcommand{\Bv}{{\boldsymbol{\mathnormal v}}}
 
 \newcommand{\Br}{{\boldsymbol{\mathnormal r}}}
 \newcommand{\Bs}{{\boldsymbol{\mathnormal s}}}

 \newcommand{\BX}{{\boldsymbol{\mathnormal X}}}
 
 \newcommand{\BM}{{\boldsymbol{\mathnormal M}}}

 \newcommand{\cF}{{\cal F}}
 \newcommand{\cG}{{\cal G}}
 
 \newcommand{\Balpha} {\ensuremath{\boldsymbol\alpha}}

 \newcommand{\Bsigma} {\ensuremath{\boldsymbol\sigma}}

 \newcommand{\Bve    }{\ensuremath{\boldsymbol\varepsilon}}

\begin{document}

\begin{frontmatter}

\title{A continuum approach to combined $\ggp$ evolution and dislocation plasticity in Nickel-based superalloys}
\author[mymainaddress]{Ronghai Wu}
\author[mymainaddress]{Michael Zaiser}

\address[mymainaddress]{Institute of Materials Simulation, Department of Materials Science, Friedrich-Alexander University of Erlangen-N\"urnberg (FAU), Dr.-Mack-Str. 77, 90762
	F\"urth, Germany}

\author[mymainaddress,TUBAF]{Stefan Sandfeld\corref{corr}}
\cortext[corr]{Corresponding author.}
\ead{stefan.sandfeld@imfd.tu-freiberg.de}
\address[TUBAF]{Chair of Micromechanical Materials Modelling (MiMM),
	Institute of Mechanics and Fluid Dynamics,
	Technische Universit\"at Bergakademie Freiberg (TUBAF),
	Lampadiusstr. 4,
	09596 Freiberg,
	Germany}

\begin{abstract}
Creep in single crystal Nickel-based superalloys has been a topic of interest since decades, and nowadays simulations are more and more able to complement experiments. In these alloys, the $\ggp$ phase microstructure  co-evolves with the system of dislocations under load, and understanding the mutual interactions is essential for understanding the resulting creep properties. Predictive modeling thus requires multiphysics frameworks capable of modeling and simulating both the phase and defect microstructures. To do so, we formulate a coupled model of phase-field evolution and continuum dislocation dynamics which adequately accounts for both statistically stored and geometrically necessary dislocations.  The simulated $\ggp$ phase microstructure with four $\gp$ variants and co-evolving dislocation microstructure is found to be in good agreement with experimental observations. The creep strain curve is obtained as a natural by-product of the microstructure evolution equations without the need for  additional parameter fitting. We perform simulations of $\gamma/\gamma^\prime$ evolution for different dislocation densities and establish the driving forces for microstructure evolution by analyzing in detail the changes in different contributions to the elastic and chemical energy density. Together with comparisons between simulated and experimental creep curves this investigation reveals the mechanisms controlling the process of directional coarsening (rafting) and demonstrates that the kinetics of rafting significantly depends on characteristics of the dislocation microstructure. In addition to rafting under constant load, we investigate the effect of changes in loading conditions and explore the possibility of improving creep properties by pre-rafting along a different loading path.
\end{abstract}


\end{frontmatter}

\section{Introduction}
Single crystal Nickel-based superalloys have been used as high temperature materials already for several decades. They essentially consist of cube-shaped precipitates of an ordered phase (termed $\g'$ phase) embedded in a face-centred cubic, solution hardened matrix (termed \g\ phase). The atomic microstructure of the coherent L1$_2$ - ordered \gp\ precipitates renders them strong obstacles to dislocation motion even at elevated temperatures, and optimizing the \gp\ fraction has played a major role in improving the thermo-mechanical properties of single crystal Ni-based superalloys \citep{Murakumo_2004_Acta}. Advances in microstructure engineering and manufacturing of single-crystal components have made this class of materials a common choice for components that require good creep resistance at high temperatures, such as turbine blades in jet engines \citep{Davis1997}. Under typical service conditions the centrifugal force within monocrystal turbine blades is acting parallel to the $\langle 001 \rangle$ crystal orientation, which is why superalloy creep (i.e. plastic deformation which occurs at load levels significantly lower than the macroscopically observed yield strength) with the stress axis oriented along a $\langle 001 \rangle$ crystal axis has been studied extensively already since the $1970$s \citep[see e.g.][]{Tien_1971_MT}. However, a significant difficulty is posed by the necessity of simultaneously investigating the time dependent aspects of plastic deformation  on the macroscopic, specimen scale as well as on the level of the dislocation and phase microstructure. Up to now this is mostly done by interrupting creep tests at distinct times or strain values, followed by characterizing the respective microstructures \citep{Wu_2016_AM, Titus_2015_AM}. However, the picture provided by even the most comprehensive experimental characterization methods is incomplete since it is impossible to directly monitor the dynamic interplay of different microstructural mechanisms and at the same time to obtain detailed information about the internal stresses which provide essential driving forces for both phase (\ggp) and dislocation microstructure evolution. This, however, is indispensable for understanding the mechanisms that control the evolution of the dislocation and phase microstructure, and that ultimately govern the mechanical material response during (creep) loading.

Simulations, on the other hand, provide full information regarding the evolution of both internal stresses and microstructure morphology on the considered scale of resolution. This may include tensorial stress and strain fields resolved on the scale of the phase microstructure, the geometry of the \gp\ precipitates, and possibly the evolution and arrangement of dislocations to name but a few. (Obviously, physical details not contained within the respective model \emph{can not} be represented, which therefore requires a careful consideration of the underlying model idealizations). The possibility to control and modify the degree of microrstructural detail 'at will' makes such simulations eminently suitable for identifying the mechanisms that govern microstructure evolution and control the microstructure-property relations.   

A number of constitutive models on the macroscopic specimen scale have been used to reproduce creep curves \citep{Oruganti_2012_AM, Kim_2016_IJP}, without explicitely considering the underlying microstructure evolution. One of the advantages of those models is that they are amenable to numerical implementation within standard finite element frameworks, which are readily available and allow for treatment of complex boundary conditions. These models can be regarded as valuable engineering tools for roughly estimating the creep life-time  \citep{Le_2014_IJP, Connolly_2014_MST}. Details of the creep curves are, however, very sensitive with respect to the specific creep test conditions  \citep{Reed_1999_AM} -- a sensitivity which is particularly pronounced in multi-stage creep processes where the microstructural mechanisms that govern the creep behavior change along the creep curve. As a consequence, not only the model parameters may need to be fitted anew for matching different experimental conditions, but even the constitutive formulation itself may need to be adjusted accordingly. While this might be a feasible approach from an engineering point of view, the predictive power of such models is necessarily  limited; such models are reliable only in the very specific situations to which they have been tailored and fitted. In particular, information regarding microstructural processes and creep mechanisms enters such models only in an indirect manner, i.e. through the selection of parameters and the choice of the constitutive model equations, and it is not clear in which precise manner the macroscopic parameters relate to the microstructural processes. This deficiency makes such models, despite their undoubted usefulness as engineering design tools, unsuitable for predictive modeling on the microstructural level and thus for microstructure engineering.  

Atomistic methods, such as the molecular dynamics (MD) simulation method are able to show in great detail, for instance, the interaction between edge dislocations and solid solution elements \citep{Zhang_2013_CMS}. MD simulations have been used to study interface dislocation networks at $\gamma/\gamma^\prime$ interfaces \citep{Prakash_2015_AM} and their interactions with matrix dislocations \citep{Zhu_2013_CMS, Wu_2011_PM}. However, owing to system size limitations, MD simulations can hardly handle large numbers of dislocations or low strain rates, and the intrinsic limitations of the method in capturing slow diffusion-controlled processes make it currently unsuitable for describing the temporal evolution of the phase microstructure. 

On the other hand, the discrete dislocation dynamics (DDD) simulation method can not only provide detailed information about dislocations, but can also deal with the evolution of large numbers of dislocations in multiple slip systems on realistic time scales. However, up to now the $\gamma^\prime$ precipitates in DDD simulations of $\ggp$ microstructures have always been assumed to be static: such simulations capture how the precipitates influence the dislocation microstructure evolution but not vice versa \citep{Probst-Hein_1999_AM, Yashiro_2006_IJP, Huang_2012_IJP, Gao_2015_JMPS}. We demonstrate in the present paper that this may provide a too limited perspective if it comes to understanding creep processes in \ggp microstructures, which are determined by the co-evolution of both phase and dislocation microstructures and cannot be reduced to dislocations moving in a static precipitate arrangment. 

The phase-field (PF) method has become one of the most popular methods for simulating phase microstructure evolution in a continuum setting. To predict the simultaneous evolution of phase and dislocation microstructure and to study their mutual interactions it is therefore natural to consider also dislocation microstructures -- or, more generally speaking, plastic deformation processes -- in a continuum setting and couple them to PF models. In this spirit, phenomenological (visco-) plasticity models have been coupled to mesoscopic PF models, and the results show that plastic activity accelerates rafting (directional coarsening of $\ggp$ structure) and causes misalignment of the $\gamma^\prime$ precipitates \citep{Gaubert_PM_2010, Cottura_2012_JMPS, Tsukada_Acta_2011, Tanimoto_2014_CMS}. The main disadvantage of these plasticity models is the lack of any information on dislocation microstructure, which limits their ability to reveal dislocation associated mechanisms. By contrast, phase-field dislocation dynamics (PFDD) offers an approach that allows to treat both phase and defect microstructure evolution in a common conceptual framework \citep{Finel_2000_MRS,Hu_2004_IJP}. In this approach, the dislocations on a given slip system are represented by a multiple-valued order parameter where each value represents a quantum of crystallographic slip (slip of a representative slip plane by one Burgers vector). Accordingly, dislocations appear as localized transitions between different values of the slip order parameters. The main problem of the method resides in its computational cost: Since the numerical grid spacing must be sufficiently small to properly resolve the dislocation core, the computational cost of PFDD may become very high, despite attempts to coarsen the PFDD length scale \citep{Rodney_2003_AM}. In fact, this approach is rather a continuum formulation of discrete dislocation dynamics, which engenders an additional computational cost because of the need to represent the dislocations by field variables even in locations where they are physically absent -- a drawback which needs to be offset against the advantage of being able to treat phase and defect microstructures within a common conceptual framework. 

An alternative continuum approach to dislocation plasticity is provided by \emph{Continuum Dislocation Dynamics} (CDD) which defines density measures to characterize the dislocation system and formulates partial differential equations to describe their evolution. This was initiated about half a century ago by \citet{Kroner_1958_book}, \citet{Nye_1953_AM} and \citet{Mura_1963_PM} who introduced a tensorial measure (the so-called \emph{Kr\"oner-Nye tensor}) for the geometrically necessary dislocation (GND) densities inside a crystal, together with the corresponding evolution equations. Up to now, a number of different models, all based on the Kr\"oner-Nye tensor, have been developed including those in \citep{Acharya_2001_JMPS, Sedlacek_2003_PM,Xia_2015_MSMSE, Le_2016_IJP}. A different line of thought distinguishes dislocations according to their direction of motion and can therefore capture the simultaneous evolution of GND and so-called 'statistically stored' dislocations of zero net Burgers vector. First steps in this direction were taken by \citet{Groma_1997_PRB, Groma_2003_AM} for a plane-strain geometry with straight edge dislocations. The approach of \citet{Groma_2003_AM} is of particular interest for simulation of $\ggp$ microstructures since the corresponding evolution equations have been shown by \citet{Yefimov_2004_JMPS} to accurately reproduce the dislocation motions and dislocation-induced internal stress and eigenstrain fields in a precipitation-hardened model material. Subsequently, Hochrainer et al. introduced a very general extension towards arbitrary configurations of curved dislocations \citep{Hochrainer_2007_PM, Stefan_2011_JMR, Hochrainer_2014_JMPS,Sandfeld_2015_IJP}, which has been successfully benchmarked by discrete dislocations dynamics (DDD) simulations \citep{Stefan_2015_MSMSE} and against a range of other continuum theories \citep{Monavari_2016_JMPS}. The attractiveness of CDD in modeling dislocation microstructure evolution resides in the fact that it allows to treat dislocation microstructures in terms of averages well above the scale of individual dislocations, and can thus achieve a very significant reduction of computational cost not only in comparison with PFDD but also with DDD.  

Coupling CDD-type plasticity models and PF models for the evolution of \ggp\ microstructures has only been demonstrated as a proof of concept for an idealized situation in \citep{Wu_2016_SM, Wu_2017_JAC}, where the present authors investigated some aspects of dislocation assisted rafting in terms of the shape evolution of a single \gp\ precipitate. Thus, collective phenomena in multiple-precipitate coarsening could not be studied. Such collective phenomena arise not only from the direct elastic interactions of multiple precipitates and the superposition of the diffusive fluxes that control precipitate evolution, but equally from the fact that precipitates mutually modulate the dislocation fluxes and dislocation-induced internal stress fields they experience. The ensuing collective phenomena may be essential for understanding the microstructure evolution. In the present work, we develop a model that couples the CDD dislocation dynamics with the PF \ggp\ evolution based on an eigenstrain formalism. Section~\ref{Phase-field model} introduces a PF model which is able to simulate realistic $\ggp$ microstructures with four $\gamma^\prime$ variants. Section~\ref{Continuum dislocation dynamics} details the dislocation density evolution. Section~\ref{Results and discussion} presents numerical results including an analysis of the influence of dislocation associated stresses and associated elastic energy contributions on multi-precipitate coarsening, a discussion of the mechanisms which control local features in the microstructure, microstructure -- creep property relations and their dependence on dislocation density, and the possibility of modifying creep properties by pre-deformation along a different loading path. We conclude with a critical discussion of the perspectives and limitations of the present approach in Section~\ref{Results and discussion}.

\section{The Phase-field model}
\label{Phase-field model}
In single crystal Nickel-based superalloys, the $\gamma^\prime$ phase has a L$1_2$ crystallographic structure of which there exist four different variants depending on the position of the Al atoms in the unit cell. Suitable PF models for such four-variant $\gamma^\prime$ microstructures can generally be divided into two different types. The first type considers ''physical`` order parameters and a realistic interface thickness which is of the order of a few nano meters \citep{Zhu_2004_AM}. Therefore, in a numerical implementation a very fine spatial  resolution, typically in the sub-nanometer regime, is required. This is also important to avoid artificial 'grid pinning' \citep{Rancourt_2016_JMPS}. This strongly limits the total size of the domain that can be computed in such a simulation. The second type of models departs from the idea that the real, physical interface is exactly represented in the PF model. There, the interface thickness is rather dictated by numerical properties. To ensure consistency with the underlying physics it is ensured that the simulated interface energy matches the physical interface energy. One of main computational advantages of this approach is that the admissible system size can be significantly extended such that realistic sizes of $\gamma/\gamma^\prime$ microstructures can be simulated. The Kim-Kim-Suzuki (KKS) model \citep{Kim_1999_PRE}, which is used throughout the present work, belongs to this type of models and has been successfully applied to single crystal Nickel-based superalloys \citep{Zhou_2010_PM}. 

We consider a Ni-Al binary system where the phase microstructure can be represented by one conserved order parameter $c$ characterizing the alloy composition in terms of the local Al concentration, and four non-conserved crystallographic order parameters $\phi_i$ $(i=1\ldots 4)$ which distinguish the four possible $\gamma^\prime$ variants. Values $\phi_{i}=0 \; \forall \; i$ characterize the $\gamma$ phase, $\phi_i=1$ for one $i$ denotes the $i$-variant of the $\gamma^\prime$ phase and a value $0<\phi_i<1$ corresponds to a $\gamma/\gamma^\prime$ interface or an interface  between different \gp variants, i.e., an antiphase boundary (APB). 
The evolution of the \ggp\ microstructure is given by equations for the concentration $c$ and the crystallographic order parameters $\phi_i$. These equations are derived under the assumption that the rates of change of $c$ and the $\phi_i$ depend linearly on the corresponding functional derivatives of a free energy functional \citep{Kim_1999_PRE}. Because the concentration is a conserved quantity and the crystallographic order parameters $\phi_i$ are non-consered quantities, the corresponding evolution equations are of Cahn-Hilliard and Allen-Cahn type, respectively. Assuming isotropic and homogeneous mobility coefficients $\text{M}$ and $\text{L}$ we thus find:
\begin{equation}
\partial_t c = M \nabla^2 \frac{\delta F}{\delta c}\quad\text{and}\quad
\partial_t \phi_i = - L \frac{\delta F}{\delta \phi_i}.
\end{equation}   
The free energy functional is assumed in the following form:
\begin{equation} \label{eq:free_energy}
F = \int_V \Big [ \cF^{\text{chem}} + \frac{K_\phi}{2} \sum^{4}_{i=1}|\nabla\phi_i|^2 + \cF^{\text{el}} \Big ] \diff V,
\end{equation} 
where $\cF^{\rm el}$ and $\cF^{\rm chem}$ are the elastic and chemical contributions to the free energy density and $\half{K_\phi} \sum^{4}_{i=1}|\nabla\phi_i|^2$ is a gradient-dependent energy term (with $K_\phi$ a gradient energy density coefficient). These terms will be detailed subsequently.

\subsection{Gradient-dependent energy density}
The gradient-dependent energy term $\half {K_\phi} \sum^{4}_{i=1}|\nabla\phi_i|^2$ in Eq. \eqref{eq:free_energy} controls the structure of the inter-phase boundaries. This contribution is treated in the spirit of phenomenological phase field modeling: Given a functional form for the chemical free energy and assuming a computational interface thickness and numerical resolution, the gradient energy parameter $K_\phi$ is adjusted such that it reproduces the correct interface energy. The interface energy can be obtained from other calculation methods such as cluster variation method \citep{Wang_2007_CMS}. The value of the parameter $K_\phi$ thus needs to be determined once the other energy contributions are specified. 

\subsection{Chemical free energy density}

The local chemical free energy density $\cF^{\text{chem}}$ governs the energy of a mixture of the $\g$ and $\gp$ phases. It is linked to the chemical free energy of the $\g$ and $\gp$ phases by:
\begin{eqnarray}
\cF^{\text{chem}} &=& \Big( 1-h\left(\allphi\right) \Big)\, \cF^\gamma 
        + h(\allphi)\,\cF^{\gamma^\prime}+ \cG(\allphi)
\end{eqnarray}
where $h(\allphi)$ is an interpolation function introduced below, and the free energies of the \g\ and \gp\ phase, $\cF^{\g}$ and $\cF^{\gp}$, are given by
\begin{eqnarray}
\cF^\g= F_0(c_\g-c_\g^e)^2\quad\text{and}\quad \cF^\gp = F_0(c_\gp-c_\gp^e)^2. 
\end{eqnarray} 
Therein, $F_0$ is the second derivative of the chemical free energy near the equilibrium composition, $c_\g$ and $c_\gp$ are the concentrations of Al in the $\g$ and $\gp$ phases, and and $c_\g^e$ and $c_\gp^e$ are the respective equilibrium concentrations. Local equilibrium requires $\diff \cF^\g / \diff c_\g = \diff \cF^\gp / \diff c_\gp$, hence $c_\g - c_\gp = c_\g^{\rm e} - c_\gp^{\rm e}$. From the Al concentrations in the components of the local $\g-\gp$ mixture, the overall concentration is evaluated with the same interpolation function $h(\allphi)$ as used for the chemical free energy:
\begin{equation}
c=[1-h(\allphi)]\,c_\g + h(\allphi)\,c_\gp.
\end{equation} 
The interpolation function $h$ is assumed to take the following form
\begin{equation}
h(\allphi) = \sum_{i=1}^{4} \big[\phi_i^3(6\phi_i^2 - 15\phi_i +10)\big].
\end{equation}
This choice ensures that $h$ fulfills the following conditions:
\begin{itemize}
	\item $h = 0$ and $\nabla_{\phi}h=0$ if $\phi_i = 0 \; \forall \; i$,
	\item $h = 1$ and $\nabla_{\phi}h=0$ if $\phi_i = 1$ and $\phi_j=0 \;\forall \; i\neq j$ (a pure variant).
\end{itemize}
%
The different crystallographic variants of the order parameter $\phi$ are distinguished by the free energy contribution 
$\cG = G_\phi g(\allphi)$ where the function $g(\allphi)$ is assumed as
\begin{equation}
g(\allphi) = \sum_{i=1}^{4} \phi_i^2(1-\phi_i^2) + \theta \sum_{i,j=1}^{4} \phi_i^2 \phi_j^2.
\end{equation}
The non-dimensional parameter $\theta$ penalizes co-existence of different variants and thus relates to the APB energy. We choose this parameter in a phenomenological manner such that the APB energy exceeds the \ggp\ phase boundary energy by a factor $>2$, thus preventing coalescence of different $\gp$ variants. Finally, the dimensional energy parameter $G_\phi$ is chosen such as to reproduce, in conjunction with
the gradient energy parameter $K_\phi$, the correct value for the $\ggp$ interface energy. 

Note that we do not consider a direct energetic coupling between the dislocation state and the crystallographic order parameters
$\phi_i$. Such a coupling is in principle provided by the APB energy (a $\g$ dislocation moving into $\gp$ trails an APB) and would 
require a more physically accurate treatment of APB effects. The implications of this simplification will be discussed in the Conclusions.  

\subsection{Elastic energy density}

The elastic free energy density plays a key role in our model because it provides a long-range coupling between different precipitates, and also between the dislocation and the phase microstructure. There are different ways of introducing the elastic energy into a phase field model. Traditionally, the phase field community states the problem in terms of Khachaturyan's elastic energy functional \citep{khachaturyan_1983_book} which derives from solving the elastic eigenstrain problem specified below by using a Green's function method to solve the stress equilibrium equation. This approach works only for infinite, elastically homogeneous systems (in simulations: for periodic boundary conditions). In the plasticity community, on the other hand, the elastic energy is normally formulated using the (plastic) eigenstrain formalism \citep{Stefan_2013_MSMSE}, and evaluations are based on the finite element methodology. Both formulations are equivalent in situations where Khachaturyan's formulation applies, however, the eigenstrain formalism has the advantage that it can be applied straightforwardly to finite systems and is readily generalized to account for elastic heterogeneity. We therefore state the elastic energy in terms of an eigenstrain formulation. 

In the following we use bold symbols for vectors or second order tensors, and non-bold symbols for scalar values. The symmetrized version of a second-rank tensor $\BX$ is denoted by $\BX^{\rm (s)}$, the symbol $\otimes$ denotes the outer (tensor) product, a dot "$\cdot$" denotes the inner product and "$:$" is the double inner product. With these notations, the elastic energy density is given by:
\begin{equation}
\cF^{\text{el}} = \frac{1}{2} \boldsymbol \sigma : \Bve^{\text{el}}
\label{eq:eedens}
\end{equation}
where the elastic strain $\Bve^{\text{el}}$ and the stress $\Bsigma$ are linked through the fourth order stiffness tensor $\IC$,
\begin{equation}
\Bsigma = \IC : \Bve^{\text{el}}.
\label{eq:hooke}
\end{equation}
In Ni-based superalloys, the $\g$ and $\gp$ phases have slightly different elastic constants. This elastic inhomogeneity may in principle affect the directional coarsening behavior \citep{Gururajan_2007_AM, Gaubert_PM_2010}. In the following we neglect the elastic inhomogeneity and use the same stiffness tensor $\IC$ for the $\g$ and $\gp$ phases. This simplification is {\em not} motivated by computational convenience -- since we use a finite element method for solving the eigenstrain problem, implementation of elastic heterogeneity would be straightforward. Our motivation rather stems from the fact that, for an elastically homogeneous microstructure, no directional coarsening occurs in the absence dislocation motion and plastic flow. Thus, the assumption of elastic homogeneity allows us to systematically elucidate the role played by dislocation motion in directional coarsening without the need to disentangle dislocation effects from superimposed effects of elastic heterogeneity.

Given that both misfit strains and typical creep strains are small, we adopt a small-strain formulation. The total strain $\Bve$ then derives from the material displacement vector $\Bu$ according to
\begin{equation} 
\Bve = (\nabla \otimes \Bu)^{\rm (s)}
\end{equation}
The strain tensor is now additively decomposed into an elastic strain $\Bve^{\text{el}}$ and an inelastic (or eigen-)strain $\Bve^{\text{inel}}$. The latter is the sum of eigenstrains due to the $\ggp$ lattice misfit, $\Bve^{\text{mis}}$, and eigenstrains caused by dislocation-mediated plastic deformation, $\Bve^{\text{dis}}$. It follows for the elastic strain tensor:
\begin{eqnarray}
\Bve^{\text{el}} &=& \Bve - \Bve^{\text{inel}} = (\nabla \otimes \Bu)^{\rm (s)}
-  (\Bve^{\text{mis}}+\Bve^{\text{dis}}).
\end{eqnarray}
The two eigenstrain tensors are given by
\begin{eqnarray}
\Bve^{\text{mis}} &=& h(\allphi) \epsilon^{\text{mis}} \matone\\
\Bve^{\text{dis}} &=& \sum\limits_{k} a^k \BM^{k},
\end{eqnarray}
where $\epsilon^{\text{mis}}$ is the volumetric eigenstrain due to $\ggp$ lattice misfit, $\matone$ is the second order unit tensor, 
$a^k$ is the scalar shear strain on the $k$-th crystallographic slip system, and the projection tensor $\BM^k$ characterizes the respective shear direction. Both $a^k$ and $\BM^k$ will be defined and discussed in the following section. 

Minimization of the elastic energy functional with respect to the displacement field $\Bu$ leads to the mechanical equilibrium equation 
\begin{equation} \label{eq:BE}
\nabla \cdot \boldsymbol \sigma = \boldsymbol 0
\end{equation} 
which has to be solved in the system domain accounting for the imposed Dirichlet or Neumann boundary conditions for the displacement field $\Bu$ in order to evaluate, via Eqs. \eqref{eq:hooke} and \eqref{eq:eedens}, the elastic energy. 

\section{The Continuum Dislocation Dynamics model}
\label{Continuum dislocation dynamics}

In the following we formulate the plastic deformation problem following the work of \citet{Groma_2003_AM} and \citet{Yefimov_2004_JMPS}. Deformation can occur by crystallographic slip on several slip systems $i$ characterized by Burgers vectors $\Bb^k$ of modulus $b$, corresponding slip vectors $\Bs^k = \Bb^k/b$, and slip plane normals $\Bn^k$ where $\Bn^k\cdot\Bb^k = 0$. We consider a two-dimensional plane strain setting where all $\Bb_k$ and $\Bn_k$ are contained in the $xy$ plane and deformation is homogeneous in $z$ direction. The plastic strain tensor can then be written as
\begin{equation}
\Bve^{\rm dis}=\sum_k \BM^k a^k\;\quad\text{with}\quad \BM^k = (\Bs^k \otimes \Bn^k)^{\rm (s)} .
\end{equation}
Owing to the homogeneity in $z$ direction, the dislocation system consists of positive and negative edge dislocations with Burgers vectors $\Bb_i$. The line direction of all dislocations is perpendicular to the $xy$ plane. We use a density-based formulations where positive and negative dislocations are represented by densities $\rho^{+,k}$ and $\rho^{-,k}$ which can simply be understood as averaged numbers of dislocations, of a given slip system $k$, per unit area in the $xz$ plane. From the sign-dependent densities $\rho^{\pm,k}$ the total dislocation density $\rho^k$ of each slip system can be calculated as $\rho^k=\rho^{+,k} + \rho^{-,k}$, the excess (signed GND) dislocation density is evaluated as $\kappa^k=\rho^{+,k} - \rho^{-,k}$, and the Kr\"oner-Nye tensor is $\Balpha^k = \kappa^k (\Bb \otimes \Be_z)$. 

\subsection{The CDD transport equations}

Dislocations move in such a manner as to reduce the elastic energy. We consider glide motion where positive and negative dislocations of slip system $k$ move along the $\Bs^k$ direction with the respective velocities $\Bv^{k,\pm} = \pm \Bs^k v^{k}_{\rm g}$. This motion produces a resolved shear strain $a^k$ with the local strain rate
\begin{equation}
\partial_t a_k(\Br,t) = (\rho^{+,k}(\Br,t)+ \rho^{-,k})v^{k}_{\rm g}(\Br,t) b.
\end{equation}
At the same time, motion of dislocations implies their spatial transport and thus a spatio-temporal evolution of the dislocation microstructure. Following \citet{Groma_2003_AM} and \citet{Yefimov_2004_JMPS} we consider the simplest possible transport model where the densities $\rho^{\pm,i}$ are considered as conserved quantities such that the dislocation density kinetics is described by the simple continuity equations
\begin{eqnarray}
\label{eq:origin rho plus evolution}
{\partial_t \rho^{+,k}}&=& \nabla (\Bv^{+,k}_{\rm g} \rho^{+,k}) = (\nabla.\Bs^k) (v^{k}_{\rm g} \rho^{+,k}),\\
\label{eq:origin rho minus evolution}
{\partial_t \rho^{-,k}}&=& \nabla (\Bv^{-,k}_{\rm g} \rho^{-,k}) = -(\nabla.\Bs^k) (v^{k}_{\rm g} \rho^{-,k}),
\end{eqnarray}
where spatial derivatives are evaluated along the respective local glide directions $\Bs^k$.

To complete the CDD framework, the dislocation velocities need to be related to the thermodynamic driving forces for dislocation glide
and possibly also to the variables characterizing the dislocation state. In principle, the thermodynamic driving force for dislocation glide on slip system $k$ is  provided by the resolved shear stress $\tau^k = \BM^k:\Bsigma$ which is work conjugate to the shear strain $a^k$. However, on the level of dislocation density evolution, the simple assumption of a proportionality between (average) resolved shear stress and dislocation velocity has long been proven inadequate for capturing the evolution of dislocation densities in confined geometries even where such a proportionality exists on the single-dislocation level \citep{Groma_2003_AM}.  Thus, more complex velocity relationships need to be formulated. 

Several approaches are available to this end: (i) Averaged velocities for dislocation densities can be obtained by direct averaging of the discrete equations of motion. This approach was proposed by \citet{Groma_1997_PRB} and elaborated by \citet{Groma_2003_AM} and recently by \citet{Valdenaire_2016_PRB}. (ii) Velocities can be derived from a free energy functional which contains statistically averaged, dislocation-density dependent stored energy contributions. Averaged elastic energy functionals for dislocation systems were recently derived by \citet{Zaiser_2015_PRB}, and in \citet{Groma_2016_PRB} it was demonstrated that velocity expressions which derive from variation of such  functionals can be matched to those obtained by direct averaging of the discrete dislocation dynamics. (iii) A more conventional method consists in simply postulating, in the spirit of phenomenological plasticity modeling, constitutive relationships between the dislocation velocities, the resolved shear stress, and the dislocation densities in such a manner as to match obervations or, on scales where direct observations are hard to come by, to reproduce behavior found in discrete dislocation dynamics simulations. This is the approach we use in the following. 

Our model is designed to reproduce dislocation behavior in situations where the motion of dislocations is over-damped and the glide velocity of an individual dislocation $i$ in slip system $k$ is proportional to the resolved shear stress acting on that dislocation, $v(\Br_i) = B \tau^k(\Br_i)$. The mobility coefficient $B$ is taken to account for the influence of the phase microstructure on dislocation motion: as experimental observations show that dislocations cannot move into $\gamma^\prime$ precipitates until the last stage of creep, whereas our  current work focuses on the early creep stages, we evaluate the effective dislocation mobility using a rule of mixtures as $B = B^\g [1-h(\allphi)] + B^\gp h(\allphi)$ where $B^\gp \approx 0$. Thus, we assume that the effective mobility of dislocations, and hence the dislocation velocity, is zero in the \gp phase. 

\subsection{Constitutive equation with Taylor-type yield stress} 
Upon averaging, the stress acting on a single dislocation is replaced by the average stress acting on dislocations in any given volume element. In performing the average, some caution is needed since the presence of correlations in the dislocation arrangement implies that the average stress controlling the dislocation velocity is {\em not} simply equal to the spatial average. Instead, a careful calculation demonstrates that the average stress acting on dislocations can be represented as a sum of several contributions \citep{Groma_2003_AM,Groma_2016_PRB,Valdenaire_2016_PRB} among which: 
(i) the local resolved shear stress $\tau^k = \BM^k:\Bsigma$; (ii) a back stress $\tau^{\text{b},k}$ which is proportional to the gradient of the excess dislocation density $\kappa^k$; (iii)  a Taylor-type yield stress $\tau^{\text{y},k}$ which acts as a "friction stress". We assume the same velocity function $v_g^{k}$ for positive and negative dislocations. It takes the form
\begin{eqnarray}
\label{eq:origin velocity stress}
v_g^k = \left
\{ 
\begin{array}{ll}
B\;\text{sign}(\tau^k + \tau^{\text{b},k})(|\tau^k + \tau^{\text{b},k}|-\tau^{\text{y},k})  &\text{if}\;\; |\tau^k + \tau^{\text{b},k}| > \tau^{\text{y},k},\\
0 & \textrm{else.}
\end{array}
\right.\label{eq:symvelo}\\
\tau^k=\boldsymbol\sigma:\mathcal{M},\quad \tau^{\text{b},k} = - \frac{DGb}{\rho}\Bs^k\cdot\nabla\kappa^k,\quad
\tau^{\text{y},k}= \upalpha \text{Gb} \sqrt{\rho}.
\end{eqnarray}  
Here, $B$ is a dislocation mobility coefficient, $G$ is the shear modulus, $\upalpha = 0.2\ldots 0.4$ and $D=0.6\ldots 1$ are two material independent, dimensionless coefficients, and the total dislocation density is given by $\rho = \sum_k (\rho^{+,k}+\rho^{-,k})$.

\section{Rafting - simulation results and discussion}
\label{Results and discussion}

\subsection{Simulation setup and numerical  methods}
We perform creep deformation simulations for a two-dimensional system deforming in plane strain, imposing periodic boundary conditions for all fields. The boundaries of the square simulation box are aligned with the axes of a Cartesian coordinate system which is rotated with respect to the cubic crystal axes in such a manner that the [10] cubic crystal axis is aligned with the diagonal of the coordinate system, $\Be_{[{10}]} = (\Be_x + \Be_y)/\sqrt{2}$, see Fig.~\ref{fig:phase_schematic}(a). Creep deformation is induced by a spatially homogeneous and temporally constant tensile stress $\sigma^{\text{ext}}=\SI{200}{MPa}$ acting along the cubic axis in $\Be_{[{10}]}$ direction. This direction is henceforth in our discussion referred to as 'parallel' direction whereas $\Be_{[{01}]}$ indicates the 'perpendicular' direction. We assume two slip equivalent slip systems with the respective slip plane normals $\Bn_I = \Be_y$ and $\Bn_{II} = \Be_x$ and slip directions $\Bs_I = \Be_x$ and $\Bs_{II} = -\Be_y$ which correspond to $[11]$ and $[1\bar{1}]$ lattice directions. The external stress $\sigma^{\text{ext}}=\SI{200}{MPa}$ leads in the two symmetrically inclined slip systems to equal resolved shear stresses of magnitude $\tau^{\text{ext}}\approx\SI{100}{MPa}$. The system is homogeneous in $z$ direction. 

The material parameters used in our simulations are typical of NiAl \ggp microstructures; elastic constants and thermodynamic parameters of the phase field model refer to a deformation temperature of $T=\SI{1253}{K}$. A compilation of all parameters is provided in Tab.~\ref{tab:table1}. 
\begin{table}
\caption{Parameters used in our simulations (partially taken from \citep{Zhou_2010_PM})}
\label{tab:table1}
\begin{tabularx}{\textwidth}{@{\extracolsep{\stretch{1}}}*{8}{c}@{}}
  \hline 
$c^e_{\gamma}$ [-] & $c^e_{\gamma^\prime}$ [-] & $\text{f}_0$ [\SI{}{J m^{-3}}] & $\text{k}_\phi$ [\SI{}{J m^{-1}}] & $\upomega$ [\SI{}{J m^{-3}}] & $\uptheta$ [-] & $\text{M}$ [\SI{}{m^{5} J^{-1}}] & $\text{L}$ [\SI{}{m^{3} J^{-1}}] \\
 $0.160$ & $0.229$ & $3.2\times10^9$  & $9.4\times10^{10}$  & $3.9\times10^{6}$  & $10$ & $1.5\times10^{-26}$ & $5.8\times10^{-9}$\\
  \hline
 $\text{D}$ [-] & $\upalpha$ [-] & $\text{C}_{11}$ [\SI{}{GPa}] & $\text{C}_{12}$ [\SI{}{GPa}] & $\text{C}_{44}$ [\SI{}{GPa}] & $\bar\epsilon^{\text{mis}}$ [-] & $\text{b}$ [\SI{}{nm}] & $\text{B}$ [\SI{}{GPa s}] \\
  $0.6$ & $0.2$ & $198$ & $138$ & $97$  & $-0.003$ & $0.25$ &  $1\times10^{-13}$\\
  \hline\\
\end{tabularx}
\end{table}

To evaluate the elastic energy and stress state, we proceed in two steps. In a first step, we solve the elastic eigenstrain problem for an infinite body without boundary tractions. Bulk behavior is mimicked by imposing periodic boundary conditions on the simulation domain. Unlike the strain-based method of evaluating the stress and elastic energy from the eigenstrain using a Green's function method, the displacement-based finite element formulation used in our simulations entails a technical subtlety: By imposing periodic Dirichlet boundary conditions for the displacements, the spatial average of the total strain is by construction set equal to zero. As a consequence, the spatial average of the elastic strain, or equivalently the internal stress, cannot be zero once inelastic eigenstrains of non-zero spatial average are present. Because of the requirement of macroscopic stress equilibrium, however, a non-zero average of the internal stress is inconsistent with the absence of boundary tractions. This problem can be resolved in a simple manner by subtracting the average eigenstrain value $\langle \boldsymbol \epsilon^{\text{inel}} \rangle = \langle \boldsymbol \epsilon^{\text{mis}}+\boldsymbol \epsilon^{\text{dis}} \rangle$ from the elastic strain in a post-processing step \citep[cf.][]{Nemat_1999_book,Sandfeld_2015_JSMTE}, i.e., we set
\begin{equation}
\Bsigma=\IC:(\boldsymbol \epsilon^{\text{el}} - \langle \boldsymbol \epsilon^{\text{inel}} \rangle).
\end{equation}
where $\epsilon^{\text{el}}$ is the elastic strain calculated from the displacement field with periodic Dirichlet conditions. 
As a result of the correction, the evaluated elastic strains and eigenstresses now have zero spatial average, as they should, and correctly represent the solution of the eigenstrain problem in an infinite body without boundary tractions. In a second step we then use the superposition principle to add the spatially constant external stress $\Bsigma^{\rm ext}$, and correct the elastic energy accordingly. 

In our simulations we use a simple cubic grid with equally-sized cubic elements and a grid spacing of \SI{20}{nm}. The model equations are non-dimensionalized by using the grid spacing to scale all lengths, using a time constant of \SI{0.7}{s} and an energy scale of \SI{1e9}{J m^{-3}}. The simulation domain is $\SI{2.56x2.56}{\micro\metre}$ in size. For solving the elastic problem \eqref{eq:BE} we use the finite element method with quadratic interpolation functions. For the phase field model as well as for the CDD problem we use the finite volume method (FVM), implemented with a first-order implicit Euler scheme. It is worth mentioning that the time scale of CDD is much smaller than the time scale of PF, because dislocations move much faster than the $\ggp$ interfaces evolve. We therefore use a staggered scheme where, between two PF time steps, dislocations may evolve for many CDD time steps until a quasi-static dislocation configuration is reached. 

The initial $\ggp$ microstructure for rafting is generated by precipitation from a supersaturated solution with $c=0.204+R$ and $\phi_i=0.25+R$ where $R$ is a small Gaussian fluctuation term. The four different $\gp$ variants are subsequently indicated by four different colors as shown in Fig. \ref{fig:phase_schematic}(b). The average dislocation density for each of the two slip systems in our simulations is $\rho_0$  in the \g\ phase and zero in the \gp phase, and we assume that initially the excess (GND) density is zero. Thus our initial conditions for the dislocation densities are generated from the initial precipitate microstructure by setting $\rho^k(\Br) = \rho_0 [1-h(\allphi)]$ and $\kappa^k(\Br)=0$.

\subsection{Creep simulation with and without dislocations}
\begin{figure}[htp] \centering
	\hbox{}\hfill
	\includegraphics[width=0.8\columnwidth]{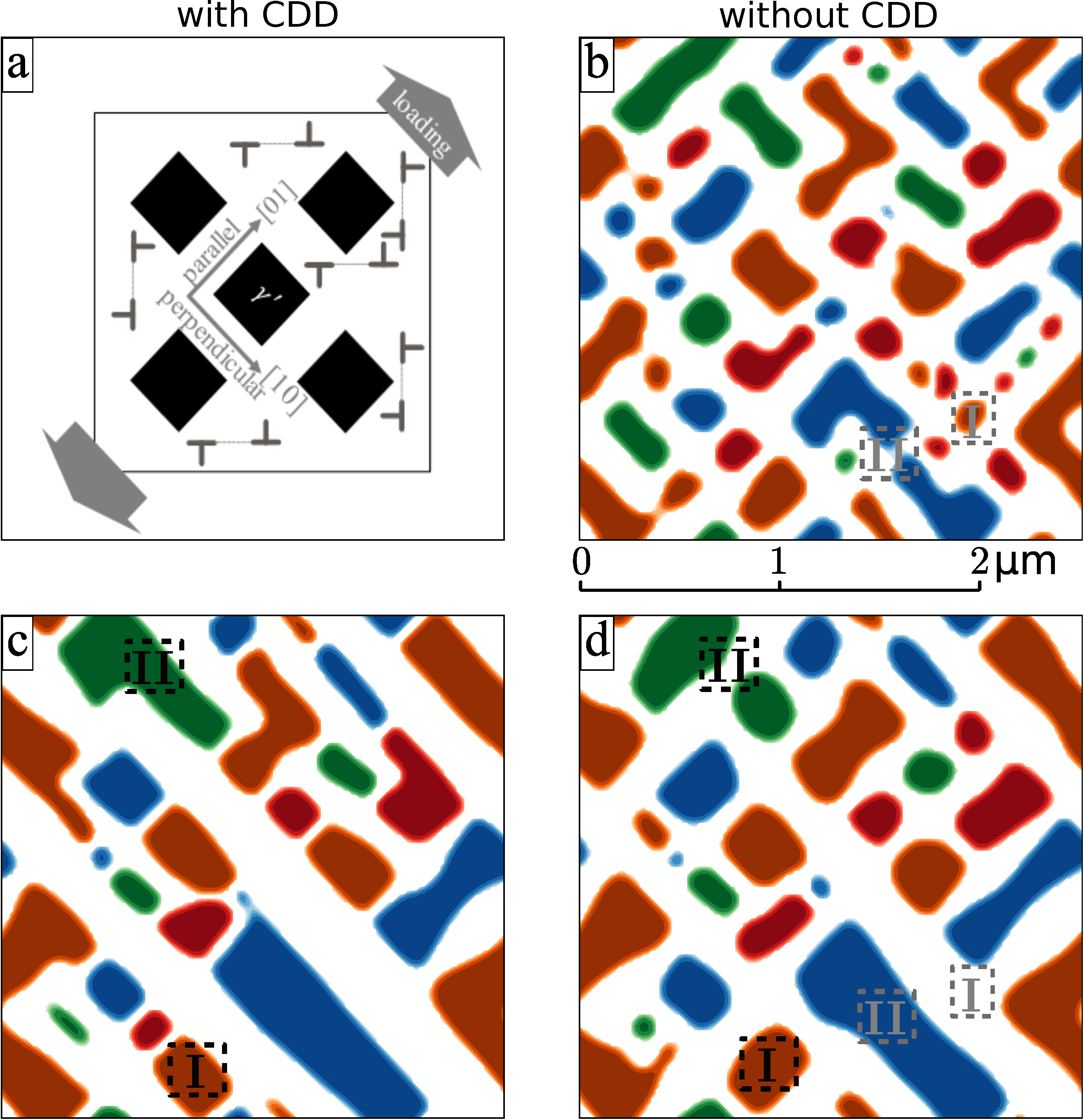}
	\hfill\hbox{}
	\caption{\label{fig:phase_schematic} Comparison of $\ggp$ morphology with and without CDD: (a) schematic representation of the system; (b) initial state; (c) intermediate state evolving under the influence of dislocation motions; (d) intermediate state without dislocations. 
		}
\end{figure}

The first set of simulations is performed without dislocations ($\rho_0 = 0$) in order to obtain reference data. As shown in Fig.~\ref{fig:phase_schematic}(b) and (d), $\gp$ precipitates are coarsening either by consuming small precipitates (grey-dashed box \rom{1}) or by merging adjacent precipitates of the same variant (grey-dashed box \rom{2}), which is energetically favorable due to the reduction of interface energy. Different $\gp$ variants, however, can not merge since the APB energy between them is higher than twice the $\ggp$ interface energy. The $\gp$ coarsening is non-directional because of the absence of elastic inhomogeneity.

A second set of simulations is performed with finite dislocation density, assuming for the dislocation density in the \g\ channels the typical value $\rho_0 = 1 \times10^{13} \SI{}{m^2}$. The simulations are started from the same phase microstructure as before and Fig.~\ref{fig:phase_schematic}(c) shows the evolved microstructure at the same time step as Fig.~\ref{fig:phase_schematic}(d). As creep proceeds, $\gp$ precipitates now preferentially coarsen in the direction perpendicular to the stress axis. This directional coarsening proceeds through two different mechanisms: (i) some precipitates immediately coarsen into the perpendicular direction, as marked by the black, dashed box \rom{1} in Fig.~\ref{fig:phase_schematic}(d). This coarsening mechanism is similar to the one that was observed by the authors for the case of a single precipitate  \citep{Wu_2016_SM, Wu_2017_JAC}. (ii) another coarsening mechanism consists in dislocations accelerating the merger of $\gp$ precipitates of the same variant which are, even at some distance, mutually aligned perpendicular to the stress axis: see the precipitates marked by black-dashed boxes \rom{2}. 

\subsection{Evolution of dislocation densities and stresses}
\begin{figure}[htp] \centering
	\hbox{}\hfill
	\includegraphics[width=\columnwidth]{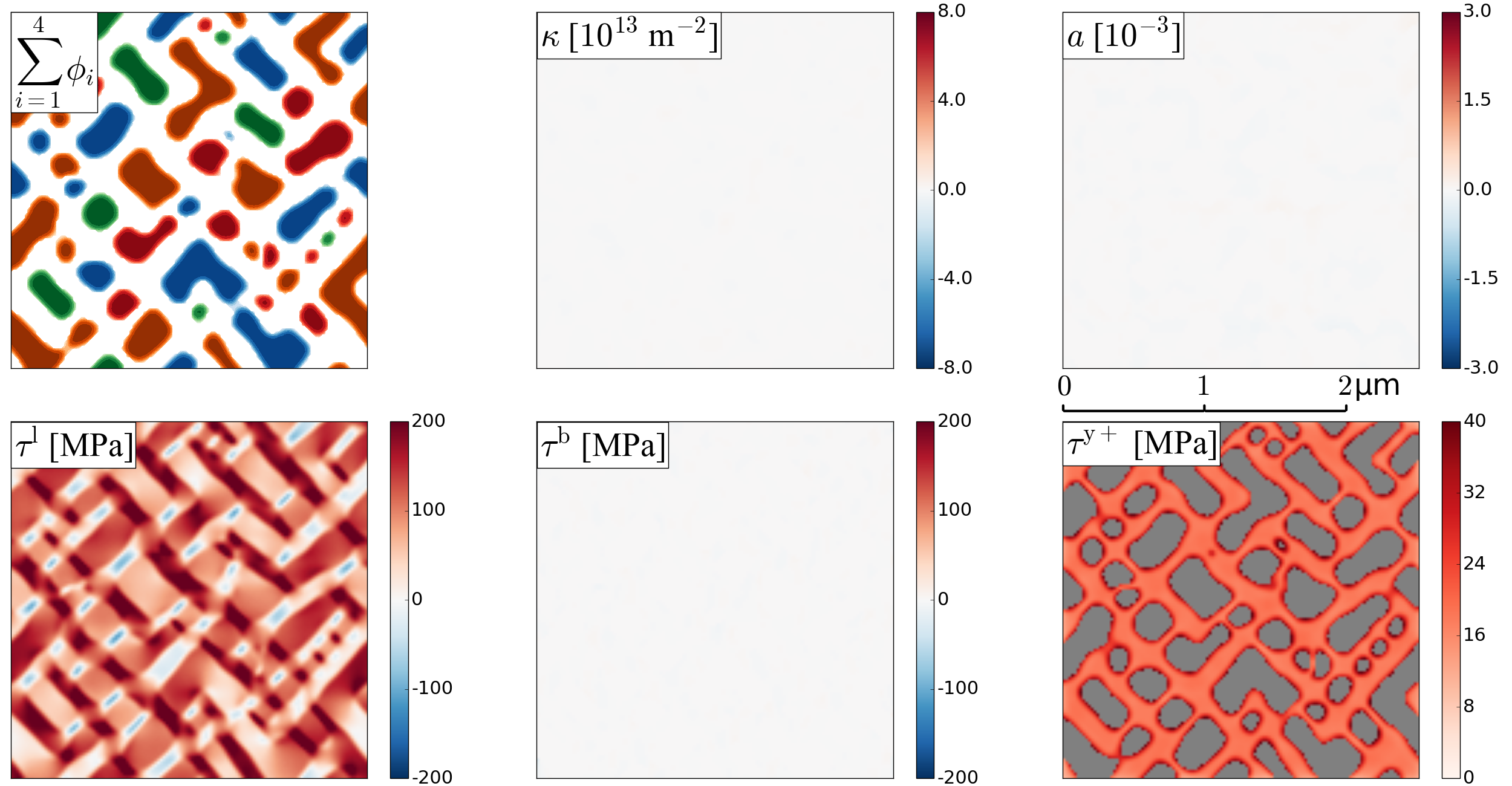}
	\hfill\hbox{}
	\caption{\label{fig:CDD_initial} Fields at initial state: (a) $\ggp$ morphology; (b) excess density; (c) plastic strain; (d) long range stress; (e) back stress; (f) yield stress.
		}
\end{figure}

\begin{figure}[htp] \centering
	\hbox{}\hfill
	\includegraphics[width=\columnwidth]{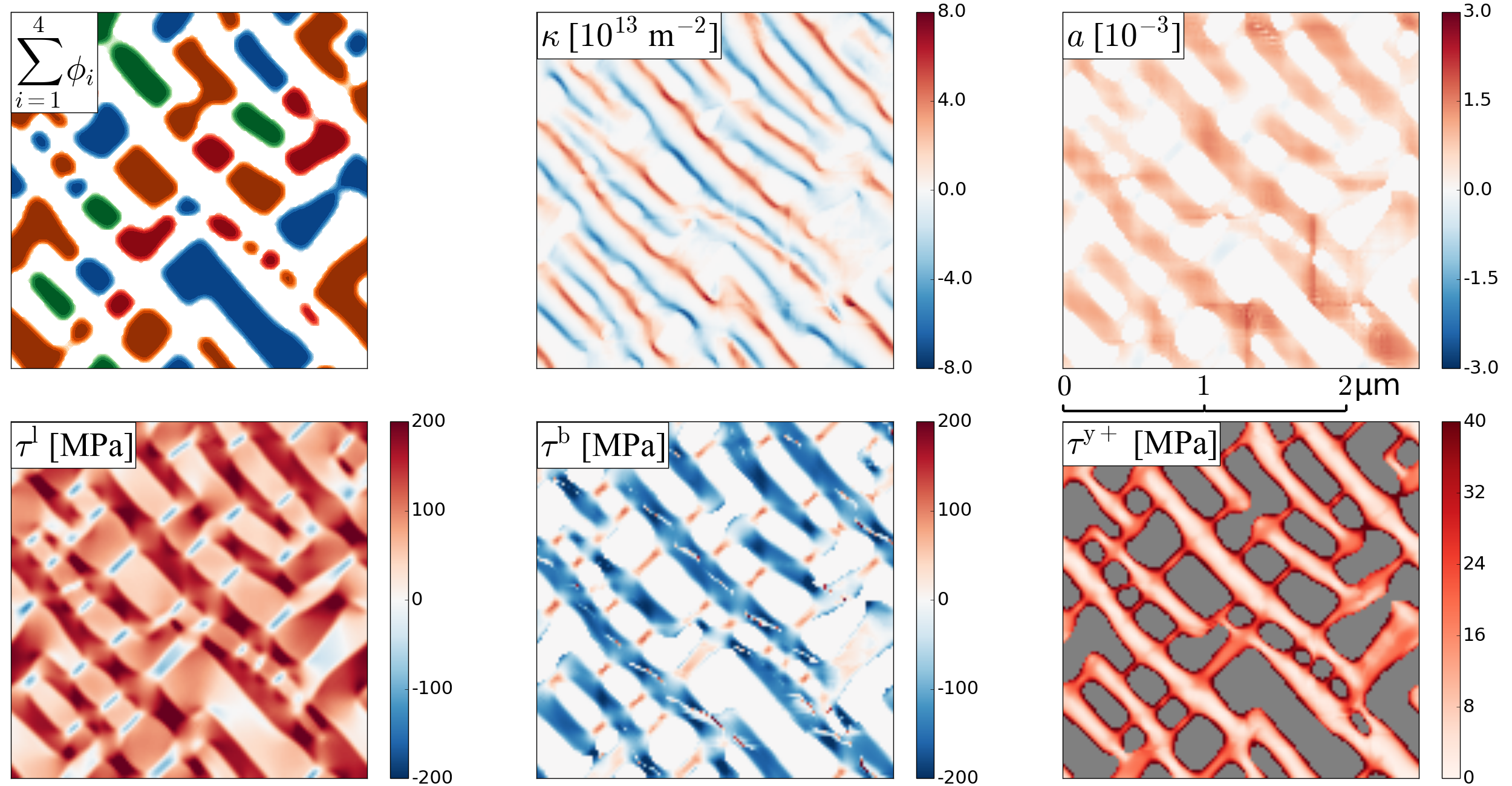}
	\hfill\hbox{}
	\caption{\label{fig:CDD_intermediate}Fields at intermediate state: (a) $\ggp$ morphology; (b) excess density; (c) plastic strain; (d) long range stress; (e) back stress; (f) yield stress.
		}
\end{figure}

To reveal why and how the dislocation arrangement evolves during creep, it is necessary to take a closer look at details of the dislocation
patterns as shown in Fig. \ref{fig:CDD_initial} at the initial time step and Fig. \ref{fig:CDD_intermediate} after \SI{1e4}{s}. The color bars are the same as in Fig. \ref{fig:CDD_initial} and Fig. \ref{fig:CDD_intermediate} in order to make the results easier comparable. Initially, only SSDs are present, the excess dislocation density is negligible, and the plastic strain is zero everywhere (see Fig. \ref{fig:CDD_initial}(b,c)). 

The long range shear stress is the sum of resolved stresses due to external loading, $\ggp$ misfit eigenstrain and dislocation eigenstrain. Initially, the dislocation eigenstrain is negligible and the stress field arises from superposition of the external stress and the stress associated with the $\ggp$ misfit. In the channels that are perpendicular to the stress axis, this stress adds to the external stress whereas, in the channels parallel to the stress axis, the contributions subtract. As a consequence, the shear  stress in the perpendicular channels is positive (\SI{200}{MPa}), while in parallel channels we find a negative value of about \SI{-40}{MPa} (see Fig. \ref{fig:CDD_initial}(d)). The initial back stress, which is proportional to the gradient of the excess dislocation density, is negligible due to the low initial excess dislocation density (see Fig. \ref{fig:CDD_initial}(e)). The initial SSD dislocation configuration results in a uniform yield stress of about \SI{40}{MPa} in the $\g$ channels (see Fig. \ref{fig:CDD_initial}(f), note that the $\gp$ precipitates, where the dislocation mobility is zero, are marked in dark grey on this plot). Since the yield stress in the perpendicular channels is below the long range shear stress, dislocations can move. In the parallel channels, by contrast, the magnitudes of the long range shear stress and the yield stress are comparable, which implies dislocations are significantly less active. Positive dislocations move in the positive slip direction under the action of a positive stress and in the negative slip direction under a negative stress, while the opposite is the case for negative dislocations. Hence, during creep, positive and negative dislocations separate in the perpendicular channels as they accumulate at the $\ggp$ interfaces, while such separation is not obvious in the parallel channels (see Fig. \ref{fig:CDD_intermediate}(b)). This agrees with experimental observations \citep{Miura_2000_Superalloy, Jacome_2013_AM}. The plastic strain distribution shows that massive dislocation motion has taken place in the perpendicular but not in the parallel $\g$ channels (see Fig. \ref{fig:CDD_intermediate}(c)). The long range shear stress is altered not only because of the $\ggp$ morphology change but also because of the changes in the dislocation configuration which correspond to the piling up of geometrically necessary dislocations at the \ggp interfaces. The dislocation pileups counter-act the joint action of external and misfit stresses as they reduce the magnitude of the long range shear stress in the perpendicular channels, while they increase its magnitude in the parallel channels, however, this effect is comparatively modest (compare Figs. \ref{fig:CDD_initial}(d) and Fig. \ref{fig:CDD_intermediate}(d)). A more important role is played by the short-range mutual repulsion of GNDs which is represented by the back stress, which acts as resistance to the formation of dislocation pile-ups and plays the main role in offsetting the long-range stresses in the perpendicular channels (see Fig. \ref{fig:CDD_intermediate}(e)). This finding corroborates the earlier conclusion of \citet{Groma_2003_AM} who demonstrated by comparison of CDD and DDD data for the deformation of a single slip channel that back stresses related to short-range GND repulsion can, even in the absence of dislocation-related long range stresses, explain the distribution of slip in constrained channels and the related size effects. Finally we note that the motion of dislocations depletes the perpendicular channels where the local flow stresses are small, whereas in the parallel channels, where little dislocation motion occurs, the yield stress remains close to its initial value and remains dominated by SSD interactions.  


\subsection{Dislocation density dependence of \ggp microstructure evolution and creep properties}   
\begin{figure}[tb] \centering
	\hbox{}\hfill
	\includegraphics[width=\columnwidth]{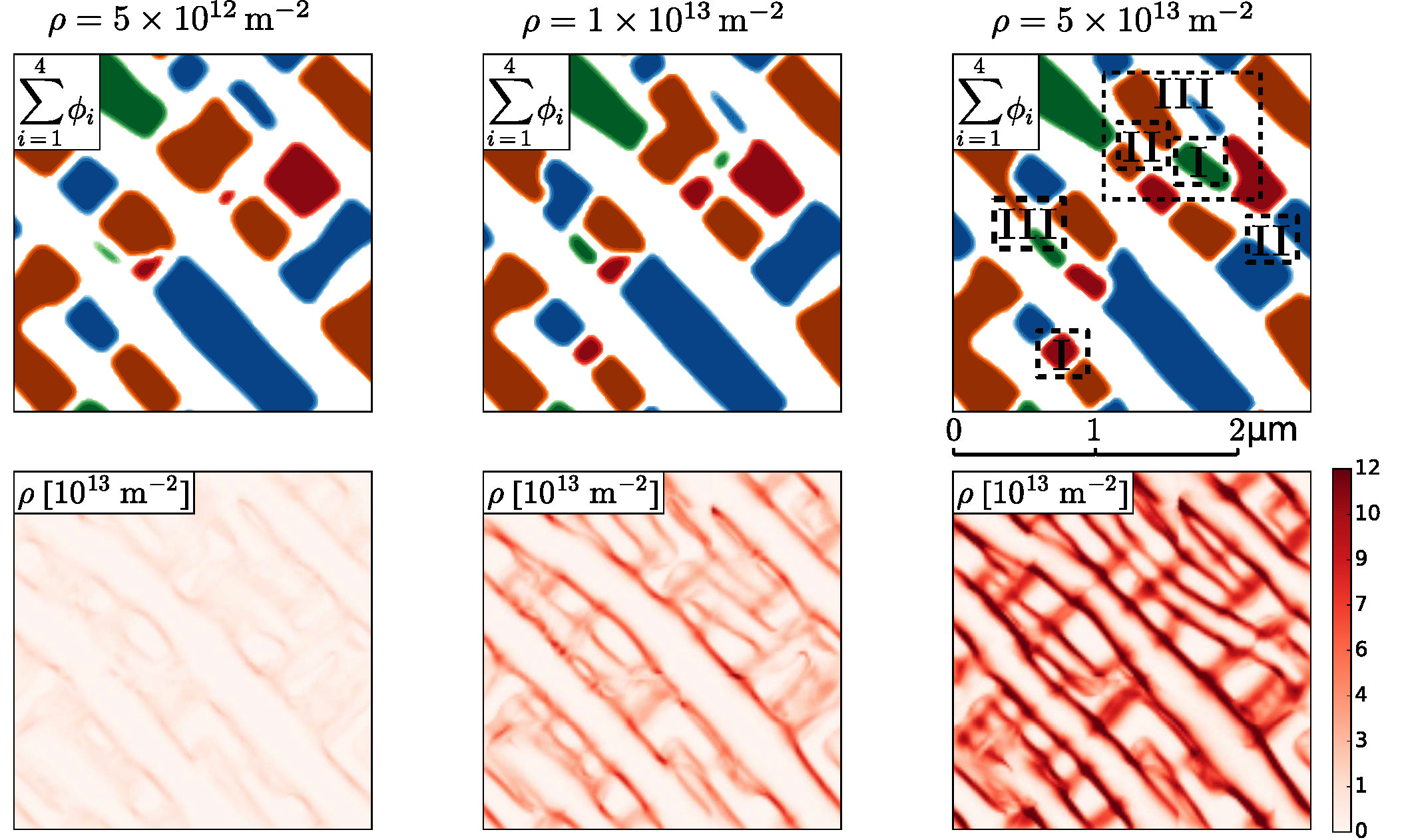}
	\hfill\hbox{}
	\caption[width=1.\columnwidth]{\label{fig:Phase_CDD}Comparison of $\ggp$ morphology (a)-(c) and total dislocation density (d)-(f) at $5.6\times10^4$ \SI{}{s} with different initial dislocation densities: (a) and (d) $5\times10^{12}$ \SI{}{m^{-2}}; (b) and (e) $1\times10^{13}$ \SI{}{m^{-2}}; (c) and (f) $5\times10^{13}$ \SI{}{m^{-2}}.
		}
\end{figure}

Accurate measurement of dislocation densities which could be used to define initial data for our simulations is still challenging. The dislocation density in single crystal Nickel-based superalloys is observed to be $5\times10^{12} \sim 5\times10^{13}$ \SI{}{m^{-2}} initially and increases to $5\times10^{13} \sim 5\times10^{14}$ at $\ggp$ interfaces when creep proceeds \citep{Miura_2000_Superalloy, Jacome_2013_AM, Nortershauser_2015_MSEA}. In 3D \ggp microstructures dislocation multiplication, leading to an increase in overall dislocation density, proceeds by expansion of loops within the \g\ channels, and consequential deposition of  GNDs at the \ggp interfaces. While this behavior can be well captured by 3D CDD simulations (see \citet{Monavari_2016_JMPS} as an example), in our 2D simulations dislocations are assumed straight and therefore dislocation multiplication cannot occur. In principle it is possible to remedy this problem by introducing dislocation sources (see e.g. \citet{Yefimov_2004_JMPS}), however, this would require us to make phenomenological and to some extent arbitrary assumptions regarding source properties. Instead, we investigate the influence of an increasing dislocation density by comparing simulations carried out for three different initial dislocation densities ($5\times10^{12}, 1\times10^{13}$ and $5\times10^{13}$ \SI{}{m^{-2}}). The $\ggp$ morphology and total dislocation density fields after $5.6\times10^4$ \SI{}{s} creep time are compared in these three cases, as shown in Fig. \ref{fig:Phase_CDD}. 

As the dislocation density increases, the resulting $\ggp$ morphologies show some gradual changes. First of all, in simulations with increased dislocation density, some small precipitates tend to be retained, such as the precipitates marked by the black-dashed boxes \rom{1} in Fig. \ref{fig:Phase_CDD}. Secondly, the presence of high dislocation densities favors the formation of connected channels in the perpendicular direction. As a consequence, precipitates which are elongated in parallel direction and block such channels, may be split in two (see black-dashed boxes \rom{2} in Fig. \ref{fig:Phase_CDD}). Moreover, while initially all $\ggp$ boundaries are basically aligned with the $[01]$ or $[10]$ orientation, local misalignment starts to appear in simulations at high dislocation density, see black-dashed boxes \rom{3} in Fig. \ref{fig:Phase_CDD}. Last but not least, the extent of rafting increases as the dislocation density increases and more dislocations pile up at $\ggp$ interfaces. Thus the bottom-line is that increasing the dislocation density promotes directional rafting but may, to some extent, counter-act coarsening. 

\begin{figure}[tb] \centering
	\hbox{}\hfill
	\includegraphics[width=\columnwidth]{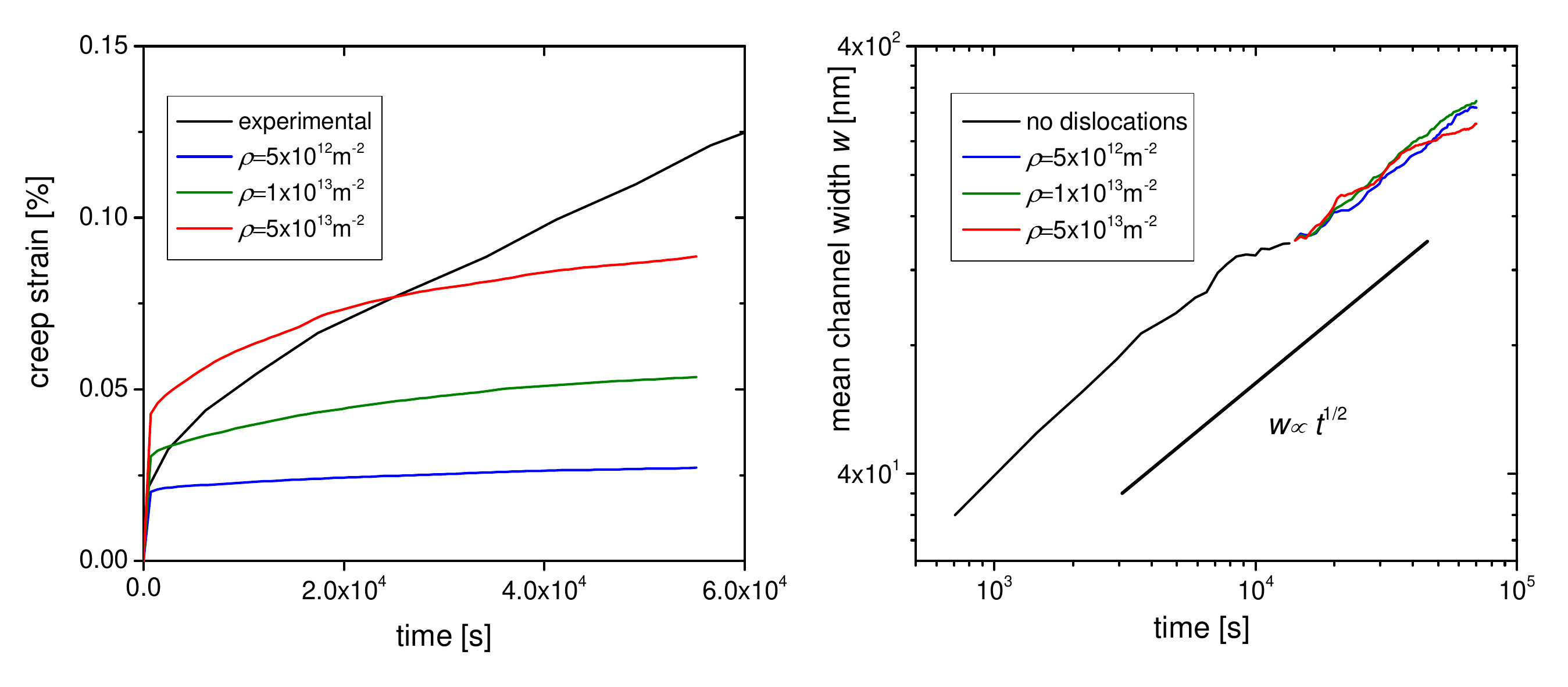}
	\hfill\hbox{}
	\caption[width=1.\columnwidth]{\label{fig:creep_coarse} Left: Creep curves for different dislocation densities and experimental creep data; right: coarsening kinetics (channel width $w$ vs. time) for different dislocation densities (coloured curves) as well as initial coarsening without dislocations (black curve), full line: $w \propto t^{1/2}$. 		}
\end{figure}

The simulated evolution of plastic creep strain with time is plotted together with experimental data at \SI{1253}{K} and \SI{200}{MPa} \citep{Link_2000_AM}. From the plots in Fig. \ref{fig:creep_coarse}, left, one can observe that the strain does not simply scale with the number of dislocations, but exhibits a non-trivial dependency on dislocation density. The creep rate in our simulations is in the first simulation steps very high, as dislocations separate and move across the \g\ channels to pile up against the \ggp\ interfaces, producing a quasi-instantaneous creep strain. This piling up leads to significant back stresses which shut down rapid dislocation flow in the \g\ channels and lead to a quasi-stationary state where further creep is controlled by the slow co-evolution of the phase and defect microstructure (motion of the \ggp\ interfaces and concomitant dislocation motions in the \g\ channels). Due to the absence of dislocation multiplication and recovery mechanisms in our model we do not reach a quasi-stationary regime of constant strain rate (secondary creep) in our simulations. Nevertheless, simulations with a dislocation density of $\rho_0 = 5\times 10^{13} \SI{}{m^{-2}}$, which is comparable to experimentally observed dislocation densities in the secondary creep regime, produce acceptable quantitative agreement with experimental data. 

While the creep process and to a lesser extent the directional coarsening morphology depend on dislocation density, the coarsening process itself (i.e., the widening of the channels and the growth of the precipitates) is practically not affected by dislocation activity. 
This process, which is  driven by interface energy reduction, follows the classical $t^{1/2}$ textbook kinetics expected for an interface energy driven process. We show this in Fig. \ref{fig:creep_coarse}, right, where we plot the mean channel width (evaluated in the direction parallel to the stress axis) as a function of time, both for the initial 'microstructure preparation' stage where dislocation activity is absent, and then for the three simulations leading to the creep curves in Fig. \ref{fig:creep_coarse}, left. It is evident that, while creep is strongly dislocation density dependent, the coarsening kinetics itself is practically unaffected by dislocation activity.

\subsection{Energetic driving forces for directional coarsening}

In order to understand how the interplay of external, misfit and dislocation related elastic fields gives rise to directional coarsening, we investigate the spatio-temporal evolution of the resulting contributions to the elastic energy. The stress can be decomposed into $\Bsigma = \Bsigma^{\text{ext}} + \Bsigma^{\text{mis}} + \Bsigma^{\text{dis}}$, referring to the externally applied stress and the respective solutions of the misfit and dislocation eigenstrain problems, and similarly the local elastic strain is given by $\Bve ^{\text{el}} = \Bve^{\text{ext}} + \Bve^{\text{mis}} + \Bve^{\text{dis}}$. Accordingly, the elastic energy density ${\cal E}^{\text{el}} =  \frac{1}{2} \Bsigma:\Bve^{\text{el}}$ can be understood as a sum of 6 contributions: 
\begin{eqnarray}
{\cal E}^{\text{el}}&=&{\cal E}^{\text{ext,ext}}+{\cal E}^{\text{ext,mis}}+{\cal E}^{\text{ext,dis}}+{\cal E}^{\text{mis,dis}}\;
+{\cal E}^{\text{mis,mis}}+{\cal E}^{\text{dis,dis}}\nonumber\\
{\cal E}^{\text{X,X}} &=& \frac{1}{2}\Bsigma^{\text{X}}:\Bve^{\text{X}},\nonumber\\
{\cal E}^{\text{X,Y}} &=& \Bsigma^{\text{X}}:\Bve^{\text{Y}}=\Bsigma^{\text{Y}}:\Bve^{\text{X}}.
\end{eqnarray}
where $\text{X,Y} \in [\text{ext,mis,dis}]$. The corresponding total (space averaged) energy contributions are denoted as $E^{\text{X,Y}}= \int {\cal E}^{\text{X,Y}}dV$. Of these, the energy contribution $\cal{E}^{\text{ext,ext}}$ is homogeneous in space and, under creep loading conditions, constant in time; it will therefore not enter our subsequent discussion. 
\begin{figure}[htp] \centering
{\large (a)}
\\[.5cm]
\includegraphics[width=\columnwidth]{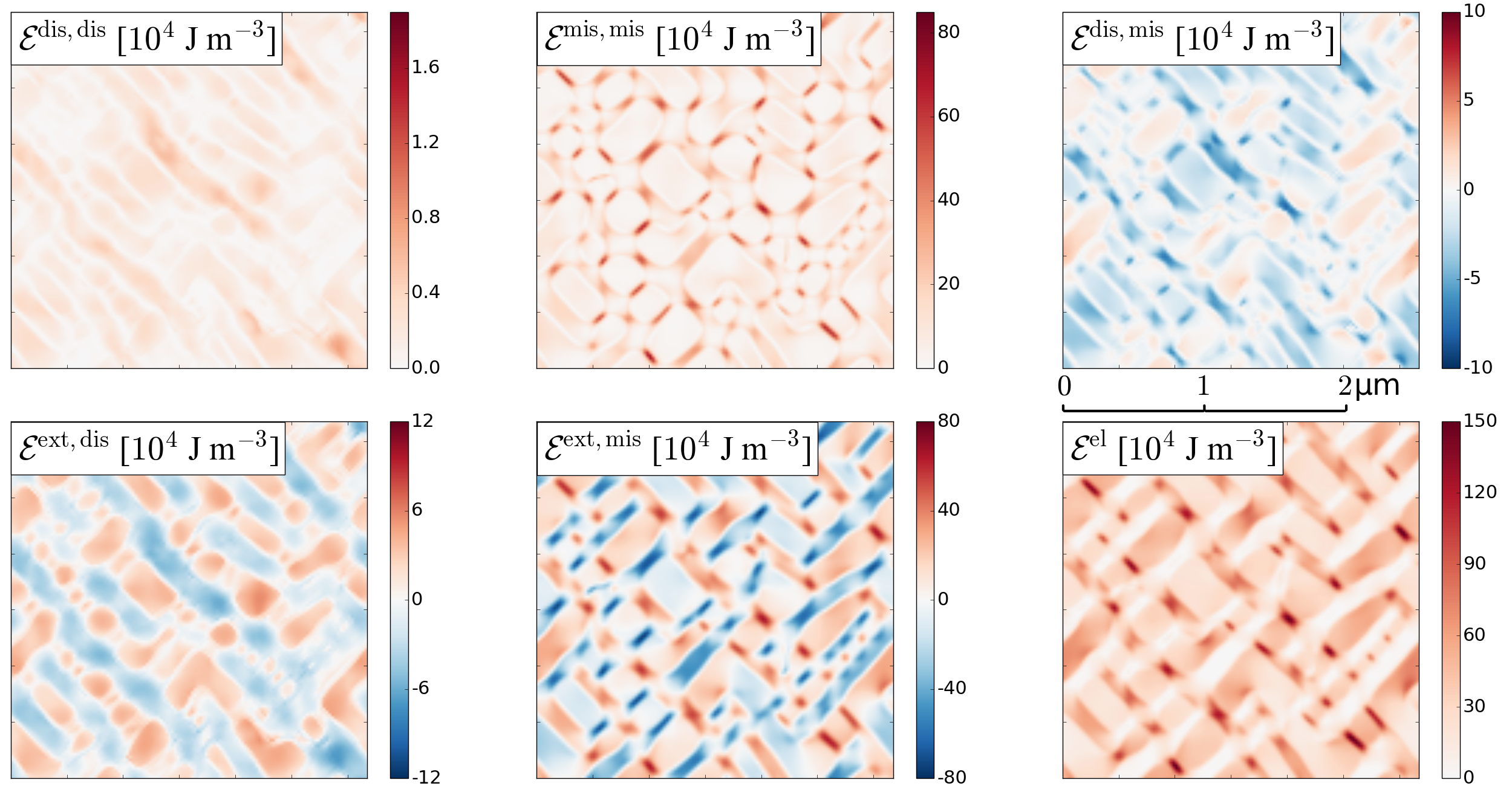}
\\[.5cm]
{\large (b)}
\\[.5cm]
\includegraphics[width=\columnwidth]{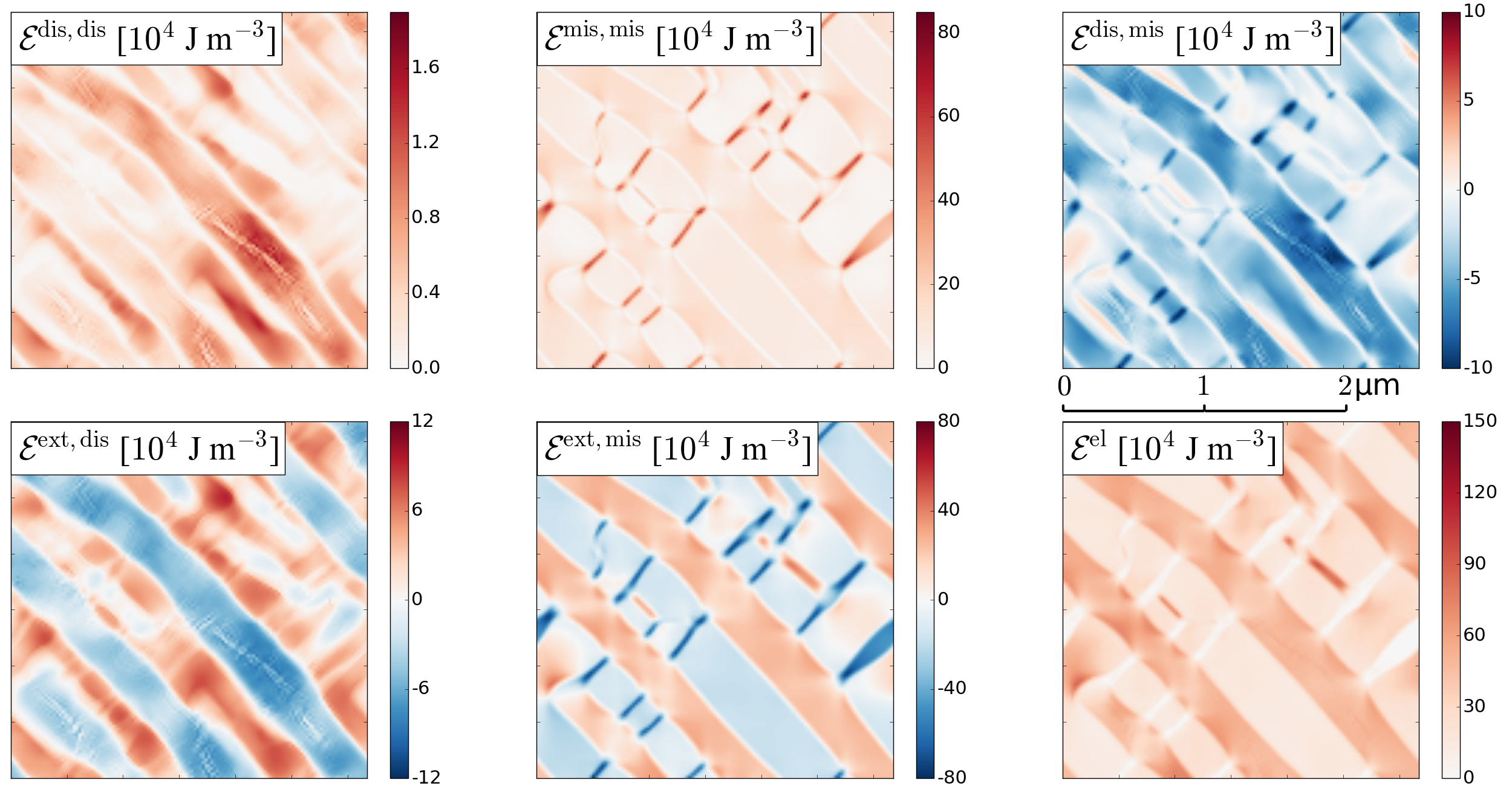}
	\caption[width=1.\columnwidth]{\label{fig:Energies} Elastic energy density contributions, (a) initial unrafted microstructure, (b) strongly rafted microstructure; in both cases energies are computed with quasi-stationary dislocation configurations under load. }
\end{figure} 
Figure \ref{fig:Energies} shows the spatial patterns formed by the different elastic energy density contributions, both in an unraftened microstructure (Fig. \ref{fig:Energies}(a)) and in a strongly rafted microstructure (Fig. \ref{fig:Energies}(b)). We make the following observations:

\begin{itemize}
\item
Dislocation related stresses are acting both in the channels perpendicular to the stress axis where they oppose the external stress, and in the precipitates as well as the parallel channels where they add to the external stress. This results in an approximately homogeneous energy contribution which increases in the course of rafting. The interaction energy between external and dislocation stresses is negative in the perpendicular channels, where dislocations relax the external stress, but positive elsewhere (first column in Fig. \ref{fig:Energies}). 
\item
Misfit related stresses lead to an energy density contribution which is everywhere positive, but somewhat lower in the precipitates than in the (vertical or horizontal channels). The misfit stresses add to the external stress in the parallel channels where this interaction further increases the elastic energy density, while the interaction reduces the energy in the precipitates and even more so in the stress-parallel channels (second column in Fig. \ref{fig:Energies}). 
\item
The interaction between misfit and dislocation related stresses leads to an energy reduction everywhere in the microstructure with the exception of the stress-parallel channels. 
\item
The total elastic energy is highest in the channels perpendicular to the stress axis, lower in the precipitates, and lowest in the channels parallel to the stress axis. This net effect results from the fact that the misfit related stresses - which produce this effect - are higher in magnitude than the dislocation related stresses which act in the opposite direction.
\end{itemize}
At first glance it might seem that these findings indicate that the system might lower its energy by coarsening in the stress-parallel direction, however, this is incorrect: The volume average of ${\cal E}^{\text{ext,mis}}$, which is the energy contribution that favors coarsening in the stress-parallel direction, is identically zero and not affected by any microstructure changes. This can easily be seen by noting that the external stress/strain is spatially constant, while the elastic stress/strain due to $\ggp$ misfit has because of stress equilibrium (Albenga's law) zero spatial average. Hence, rafting cannot be attributed to the misfit energy asymmetry between channels that are parallel and perpendicular to the stress axis. We note that the same argument applies to the interaction energy ${\cal E}^{\text{ext,dis}}$. The misfit energy $\cal E^{\text{mis,mis}}$ may change but does not distinguish between parallel and perpendicular channels. This leaves a reduction of ${\cal E}^{\text{dis,mis}}$ as the only potential driving force for directional coarsening.

\begin{figure}[htp] \centering
	\hbox{}\hfill
	\includegraphics[width=0.7\columnwidth]{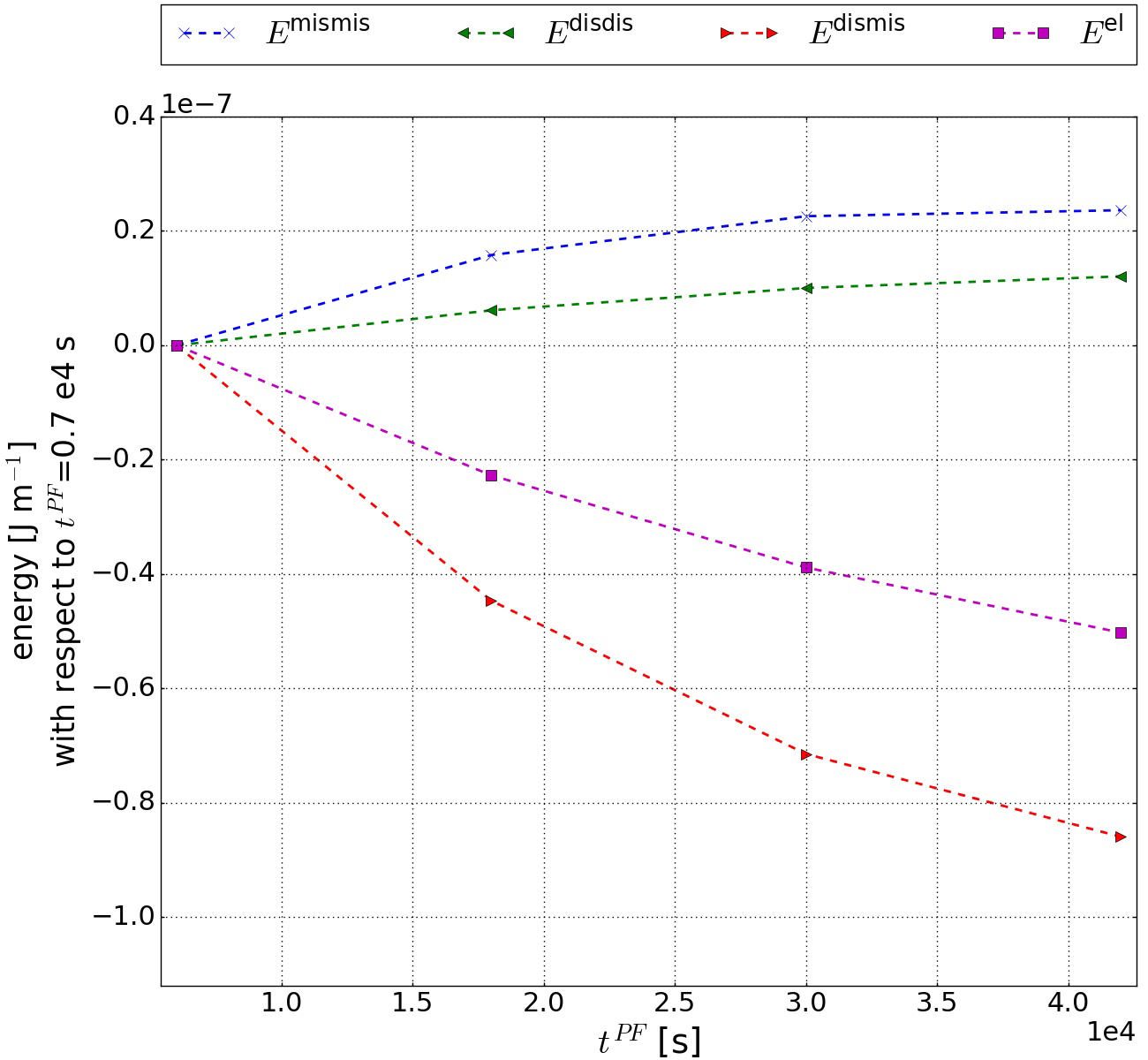}
	\hfill\hbox{}
	\caption[width=\columnwidth]{\label{fig:Energy_time} 
	Time evolution of different elastic energy contributions during rafting. The energy contributions $\cal E^{\text{ext,mis}}$ and 
	$\cal E^{\text{mis,mis}}$ are constant in time and therefore not included in the figure.}
\end{figure}

Fig. \ref{fig:Energy_time} shows the changes of the different elastic energy contributions during rafting, taking the initial unrafted microstructure as the starting point. In line with the above argument, both $\cal E^{\text{ext,dis}}$ and $\cal E^{\text{ext,mis}}$ do not change with time. The misfit energy $\cal E^{\text{mis,mis}}$ increases during directional coarsening, indicating that a rafted microstructure provides a less efficient accommodation of misfit stresses if compared to an unrafted one. To understand what provides the driving force for rafting we need to look at the dislocation-related stresses which partly relax the misfit stress in the perpendicular channels, leading to a strong decrease of the corresponding interaction energy which leads to a decrease of the total elastic energy. Thus, we conclude that directional coarsening occurs because the breaking of the symmetry between stress-parallel and stress-perpendicular channels leads to a more efficient plastic relaxation of misfit stresses as compared to a phase microstructure which retains the original cubic symmetry. 

\subsection{Microstructure evolution and creep behavior of pre-rafted \ggp microstructures}

Rafting has a negative effect on creep properties since the rafting process widens -- and thereby softens -- the channels perpendicular to the tensile stress axis where plastic activity is concentrated. This observation has led to the idea that it might be beneficial to pre-raft  \ggp microstructures by applying an initial compressive stress, thus reversing the shear stresses and interchanging the role of channels parallel
and perpendicular to the stress axis \citep{Mughrabi_2000_AEM}. Such pre-rafting leads to shrinkage of the channels perpendicular and widening of the channels parallel to the stress axis. If a pre-rafted microstructure is then deformed in tension, the active (stress-perpendicular) channels are initially narrow and thus strong, and one may hope that this has beneficial consequences for the creep properties. 

\begin{figure}[htp] \centering
	\hbox{}\hfill
	\includegraphics[width=0.7\columnwidth]{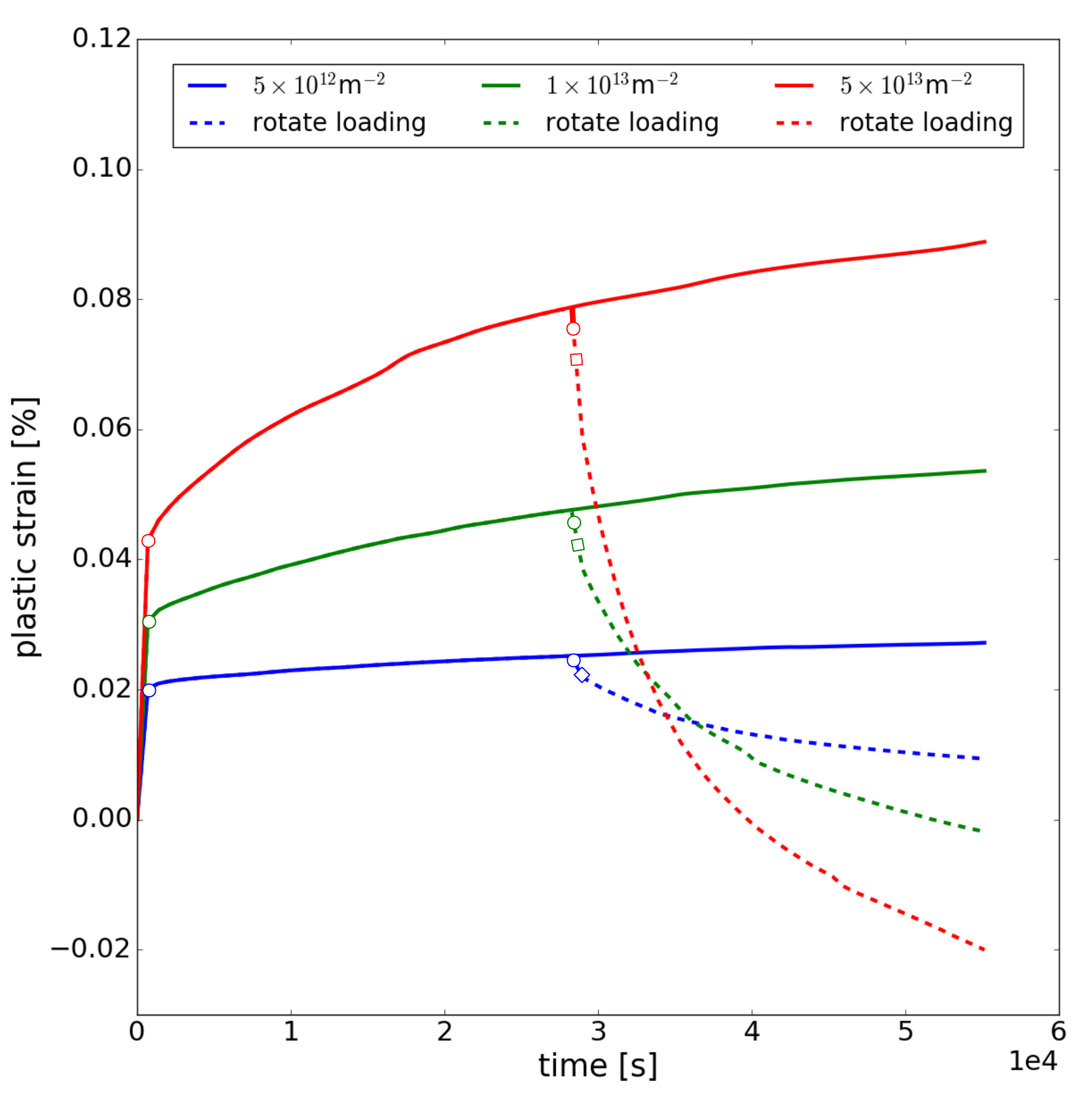}
	\hfill\hbox{}
	\caption{\label{fig:creep_rotate} Creep curves for different dislocation densities with change in deformation path 
	(stress axis rotation by 90 degrees) at $t=28000$ s. Full lines: original stress axis, dashed lines: rotated stress axis. 
	Open circles: quasi-instantaneous strain after loading and unloading along original stress axis; open squares: quasi-instantaneous 
	strain after re-loading along rotated stress axis.}
\end{figure}

To assess the consequences of pre-rafting in our model, we interchange the roles of stress-parallel and stress-perpendicular channels by simply
rotating the stress axis by 90 degrees, halfway into a simulation. In our slip geometry this has exactly the same consequences as a reversal in sign of the axial stress: the resolved shear stresses in the $\g$ channels and hence the direction of creep are reversed. The consequences for the creep behavior are shown in Fig. \ref{fig:creep_rotate}: Upon unloading, we see a small reverse strain caused by the back stresses present in the microstructure. Then, upon re-loading along the new stress axis, we observe a small, quasi-instantaneous deformation as dislocations move under the reversed shear stress. However, in the pre-rafted \ggp\ microstructure this strain is much smaller than the corresponding quasi-instantaneous strain during initial loading of the un-rafted \ggp\ microstructure (compare open circles and open squares in Fig. \ref{fig:creep_rotate}). The reason is obvious: The channels active after stress axis rotation are narrow, and consequentially deformation of these channels induces high back stresses which rapidly cause dislocation activity to stop. In this sense the pre-rafted microstructure fulfills its promise. However, the picture changes as soon as one enters the subsequent stage of co-evolution of the dislocation and phase microstructure: In this stage the rates of the 'backward' creep in the pre-rafted microstructure are much higher than those during the initial, forward creep deformation. This effect is present for all dislocation densities investigated but becomes most pronounced at the highest dislocation density
where, despite the small quasi-instantaneous deformation, the reverse strain at the end of the simulation exceeds the deformation accumulated during pre-straining.  

Thus we can conclude that the short-term consequences of pre-rafting are beneficial: We find a strongly reduced quasi-instantaneous strain during the initial, dislocation-controlled deformation stage. However, the subsequent creep behavior which is governed by the co-evolution of dislocation and phase microstructure is adversely affected. Because of the mismatch between stress state and microstructure orientation, the thermodynamic driving forces for reconstruction of the phase microstructure are much increased and accordingly this process is significantly accelerated. However, the reconstruction of the phase microstructure goes along with dislocation motions (more so if more dislocations are present) and these motions entail a significantly increased creep activity. Thus, pre-rafting can be envisaged as a double-edged method which, while providing benefits in the short term, may on longer time scales lead to a significant deterioration in creep properties. This observation emphasizes the importance of envisaging dislocation behavior and dislocation creep strain in \ggp microstructures not in isolation  but always in conjunction with the evolution of the phase microstructure.

\section{Discussion of the model}

Our model captures essential features of the rafting process driven by the interplay of externally applied stresses, misfit stresses and dislocation related stresses. In particular it demonstrates that no directional coarsening is possible in absence of dislocation activity, which leads to an elastic energy asymmetry between channels parallel and perpendicular to the stress axis that provides the thermodynamic driving force for directional, rather than isotropic coarsening. In this sense we claim that the present model provides us with a minimal model of directional coarsening driven by dislocation plasticity. We also note that the model correctly captures the piling up of dislocations against the phase boundaries. The concomitant forward and back stresses are  formally expressed in terms of gradients of the geometrically necessary dislocation density and thus of second-order strain gradients: they introduce an internal length scale into the model (the dislocation spacing) which endows the model with the capability of describing size effects: In our model, narrow channels are stronger than wide channels. The same terms also imply stress concentrations at \ggp boundaries which may be essential for understanding why, in later creep stages, dislocations may be able to penetrate the precipitates. Conventional crystal plasticity models without dislocation transport and dislocation-related internal length scales cannot capture these phenomena.

Nevertheless there is ample scope for improving the model. Maybe most conspicuous is the observation that both dislocation multiplication and dislocation recovery are missing from the present formulation. In plane strain simulations considering only edge dislocations, there is strictly speaking no mechanism for multiplication, however, it is common to introduce multiplication through phenomenological ad-hoc rules which are supposed to mimic the action of dislocation sources, see e.g. \citet{Yefimov_2004_JMPS} or \citet{Quek_2013_MSMSE}. However, dislocation multiplication in \ggp\ microstructures occurs not by sources popping out loops, but rather by moving dislocations  expanding in the \g\ channels and depositing dislocations at the \ggp\ interfaces. Since this process is intimately connected to the motion of dislocations in the heterogeneous phase microstructure, the introduction of dislocation sources through ad-hoc rules may not adequately capture the relevant physical processes and we have refrained from doing so. A more suitable approach, which may provide a both straightforward and promising generalization of the present model, would be to adopt a 3D CDD (or, for the present plane strain problem, a 2.5D CDD) formulation of dislocation kinematics as demonstrated by \citet{Monavari_2016_JMPS} for the general problem of channel slip. More challenging is the inclusion of recovery processes, since recovery in \ggp\ microstructures occurs by climb motion which allows the deposited dislocations to move along the $\ggp$ interfaces and ultimately meet recombination partners of opposite sign. 

Dislocation recovery by climb leads to a removal of interface dislocations from the system and therefore delimits the internal stresses that build up in the \g\ channels - in other words, the recovery of dislocations can here lead to a recovery of internal stresses. As a consequence of the interplay between dislocation multiplication, piling up and recovery a dynamic equilibrium might be reached which corresponds to a near-constant creep rate, i.e., secondary creep. The fact that multiplication and recovery are not included restricts our model to the primary creep stage and conversely, an extension to secondary creep will require the incorporation of climb and recovery processes into the model. However, a physically-sound description of dislocation climb is non-trivial: the climb force includes a mechanical contribution from the normal component of the long range elastic stress tensor as well as osmotic contributions associated with local vacancy equilibrium \citep{Haghighat_2013_AM, Geslin_2014_APL, Geslin_2015_PRL}. Similar to the phase microstructure evolution, dislocation climb processes are controlled by diffusion, hence climb and directional coarsening may occur at comparable rates.  We will introduce and systematically investigate the role of dislocation climb in our future work in order to enhance the physical predictivity of our multiphysics model and extend its applicability to higher creep stages. 

We finally comment on the description of the interactions between phase and defect microstructure in our model. We consider two types of interaction - an elastic interaction due to the superposition of external, misfit and dislocation induced elastic fields, and a kinematic interaction which accounts for the impenetrability of \gp\ precipitates to dislocations in terms of a reduced (zero) mobility. The latter description has the consequence that, while a \g\ dislocation cannot move into the \gp\ phase, a moving \ggp\ phase boundary can move over a \g\ dislocation which is then incorporated into the \gp\ precipitate as a pinned dislocation. This is physically not entirely realistic, since incorporation of an ordinary dislocation into a \gp variant necessarily implies creation of an antiphase boundary (APB) and the APB energy provides a very significant driving force which tries to push the dislocation out of the precipitate. This direct energetic coupling between the dislocation fields and the \gp order parameters is at present not included in our model. It might however be essential to include this aspect if one wants to describe tertiary creep, where dislocations overcome the repulsion due to APB energy and penetrate the precipitates, leading to precipitate shearing, slip localization and failure. 

\section{Summary}
We have developed a multiphysics model coupling a phase-field and a continuum dislocation dynamics model, and have applied this model to study high temperature and low stress creep in single crystal Nickel-based superalloys. The following conclusions can be drawn from the present work: 
\begin{enumerate}
\item Directional coarsening is driven by the interplay of external stresses, misfit stresses and dislocation-induced stresses. The basic mechanism can be understood as follows: The superposition of misfit and external stress favors slip in the \g\ channels that are perpendicular but not in those that are parallel to the stress axis. As soon as dislocation activity sets in, the motion of dislocations leads to a relaxation of misfit stresses and hence to a reduction of the elastic energy in the perpendicular channels (note that in the absence of an asymmetry in dislocation activity there is no such energy asymmetry). Widening the perpendicular channels renders this mechanism more efficient and consequentially, the perpendicular channels widen whereas the parallel channels shrink: We see directional coarsening. The present 2D dislocation dynamics model provides a minimal dislocation based theory to capture this phenomenon. 
\item  Both creep microstructure and creep behavior show significant dislocation density dependence: Higher dislocation density accelerates both rafting and creep deformation.
\item Dislocation and phase microstructure evolution need to be considered in conjunction. Changes in phase microstructure that inhibit dislocation motion in the short term (pre-rafting) may lead to accelerated phase microstructure changes and increased dislocation activity (hence, deterioration in creep resistance) in the long term. 
\end{enumerate}
The present model captures the basic mechanisms of rafting and provides an adequate description of the first creep stage. Improvements are still needed to provide a correct  description of three-stage creep curves. These improvements concern the inclusion of dislocation multiplication and recovery into the dislocation dynamics, such as to capture the dynamic equilibrium of both processes which leads to a quasi-stationary behavior in secondary creep. Furthermore, an improved description of the interaction between dislocations and the crystallographic order parameters of the \gp\ phase is needed to correctly describe the interaction of dislocations and precipitates, in particular the possibility for dislocations to penetrate the precipitates at high stresses/high dislocation densities, which is responsible for softening and failure in creep stage III. 
\section*{Acknowledgment}
Financial support from Deutsche Forschungsgemeinschaft (DFG) through Research Unit FOR1650 'Dislocation-based Plasticity' (DFG grants SA2292/1-2 and ZA171/7-1 ) is gratefully acknowledged.

\section*{References}
\bibliography{literature}

\end{document}